\documentclass[12pt,a4paper]{article}

\usepackage{amsmath,amssymb}
\usepackage{geometry}
\usepackage{setspace}
\usepackage{hyperref}
\usepackage{graphicx}
\usepackage{authblk}

\title {A new class of degenerate solutions to the massless Dirac equation and their potential applications in optical memories}
\author[1]{Georgios N. Tsigaridas\footnote{E-mail: gtsig@mail.ntua.gr}}
\author[2]{Aristides I. Kechriniotis}
\author[2]{Christos A. Tsonos}
\author[3]{Konstantinos K. Delibasis}
\affil[1]{Department of Physics, School of Applied Mathematical and Physical Sciences, National Technical University of Athens, GR-15772 Zografou Athens, Greece}
\affil[2]{Department of Physics, University of Thessaly, GR-35100 Lamia, Greece}
\affil[3]{Department of Computer Science and Biomedical Informatics, University of Thessaly, GR-35131 Lamia, Greece}
\date{31 March 2026}
\begin{document}
\maketitle

\begin{abstract}
In this article, we present a novel class of degenerate solutions to the massless Dirac equation, corresponding to a wide variety of electromagnetic 4-potentials and fields, including both zero field and circularly polarized electromagnetic waves.  An interesting property of these solutions is that the spin of the particles rotates in synchronization with the electric and magnetic fields of the electromagnetic waves. These results could be utilized for the development of optical memories based on materials supporting massless Dirac fermions, such as graphene.  
\end{abstract}


In \cite{1} we have shown that all solutions to the Dirac equation  
\begin{equation}\label{1}
i{\gamma ^\mu }{\partial _\mu }\Psi  + {a_\mu }{\gamma ^\mu }\Psi  - m\Psi  = 0
\end{equation}
satisfying the conditions ${\Psi ^\dag }\gamma {\rm{ }}\Psi  = 0$  and ${\Psi ^T}{\gamma ^2}{\rm{ }}\Psi  \ne 0$  are degenerate, corresponding to an infinite number of electromagnetic 4-potentials given by the formula
\begin{equation}\label{2}
{b_\mu } = {a_\mu } + {\theta _\mu }s
\end{equation}
where  $s$ is an arbitrary real function of spatial coordinates and time and 
\begin{equation}\label{3}
\left( {{\theta _0},{\theta _1},{\theta _2},{\theta _3}} \right) = \left( {1, - \frac{{{\Psi ^T}{\gamma ^0}{\gamma ^1}{\gamma ^2}\Psi }}{{{\Psi ^T}{\gamma ^2}\Psi }}, - \frac{{{\Psi ^T}{\gamma ^0}\Psi }}{{{\Psi ^T}{\gamma ^2}\Psi }},\frac{{{\Psi ^T}{\gamma ^0}{\gamma ^2}{\gamma ^3}\Psi }}{{{\Psi ^T}{\gamma ^2}\Psi }}} \right)
\end{equation}
Here, ${\gamma ^\mu }$  are the four contravariant gamma matrices in the Dirac representation 
\begin{equation}\label{4}
{\gamma ^0} = \left( {\begin{array}{*{20}{c}}
{{\sigma ^0}}&0\\
0&{ - {\sigma ^0}}
\end{array}} \right){\rm{,  }}{\gamma ^\mu } = \left( {\begin{array}{*{20}{c}}
0&{{\sigma ^\mu }}\\
{ - {\sigma ^\mu }}&0
\end{array}} \right),{\rm{  }}\mu  = 1,2,3
\end{equation}
where ${\sigma ^\mu }$ are the standard Pauli matrices given by the following formulae \cite{2}
\begin{equation}\label{5}
{\sigma ^0} = \left( {\begin{array}{*{20}{c}}
1&0\\
0&1
\end{array}} \right){\rm{    }}{\sigma ^1} = \left( {\begin{array}{*{20}{c}}
0&1\\
1&0
\end{array}} \right){\rm{   }}{\sigma ^2} = \left( {\begin{array}{*{20}{c}}
0&{ - i}\\
i&0
\end{array}} \right){\rm{  }}{\sigma ^3} = \left( {\begin{array}{*{20}{c}}
1&0\\
0&{ - 1}
\end{array}} \right)
\end{equation}
The matrix $\gamma $  is defined as $\gamma  = {\gamma ^0} + {\gamma ^0}{\gamma ^5}$, where ${\gamma ^5} = i{\gamma ^0}{\gamma ^1}{\gamma ^2}{\gamma ^3}$. Furthermore, $m$ is the mass of the particles and ${a_\mu } = q{A_\mu }$ , where $q$ is the charge of the particles and ${A_\mu }$  the electromagnetic 4-potential. It should also be mentioned that the Dirac equation is expressed in natural units, where $\hbar  = c = 1$. 

In \cite{1} we have also shown that the general form of the degenerate spinors is the following
\begin{equation}\label{6}
\Psi  = u\left( {\begin{array}{*{20}{c}}
{\bar w}\\
1\\
{\bar w}\\
1
\end{array}} \right) + v\left( {\begin{array}{*{20}{c}}
1\\
{ - w}\\
{ - 1}\\
w
\end{array}} \right)
\end{equation}
where $u,v,w$ are arbitrary complex functions of the spatial coordinates and time and  $\bar w$ is the complex conjugate of $w$. 

Based on these results we have found that all spinors of the form 
\begin{equation}\label{7}
\Psi  = \left( {\begin{array}{*{20}{c}}
{{c_1}{e^{i\left( {{\varphi _1} - h} \right)}}f + {c_2}{e^{i{\varphi _2}}}}\\
{ - {c_2}{e^{i\left( {{\varphi _2} + h} \right)}}f + {c_1}{e^{i{\varphi _1}}}}\\
{{c_1}{e^{i\left( {{\varphi _1} - h} \right)}}f - {c_2}{e^{i{\varphi _2}}}}\\
{{c_2}{e^{i\left( {{\varphi _2} + h} \right)}}f + {c_1}{e^{i{\varphi _1}}}}
\end{array}} \right)
\end{equation}
where ${c_1},{c_2},{\varphi _1},{\varphi _2}$ are arbitrary real constants and $f,h$ are arbitrary real functions of $z - t$, are degenerate solutions to the massless Dirac equation corresponding to the following 4-potentials
\begin{equation}\label{8}
\left( {{b_0},{b_1},{b_2},{b_3}} \right) = \left( {1, - \frac{{2\cos \left( h \right)f}}{{1 + {f^2}}}, - \frac{{2\sin \left( h \right)f}}{{1 + {f^2}}}, - 1 + \frac{2}{{1 + {f^2}}}} \right)s
\end{equation}
Obviously, the function $f$ must be appropriately chosen so that the spinor $\Psi$ is square integrable. Using the formulae \cite{3}
\begin{equation}\label{9}
{\bf{E}} =  - \nabla U - \frac{{\partial {\bf{A}}}}{{\partial t}},{\rm{ }}{\bf{B}} = \nabla  \times {\bf{A}}
\end{equation}
we can calculate the electromagnetic fields corresponding to the 4-potentials given by Eq. (8). Specifically, we have found that
\begin{equation}\label{10}
\begin{array}{c}
{\bf{E}} = \frac{1}{q}\left( {\frac{{2\cos \left( h \right)\left( { - s\left( {{f^2} - 1} \right)f' - f\left( {{f^2} + 1} \right){s_t}} \right) - 2f\left( {{f^2} + 1} \right)h'\sin (h)s}}{{{{\left( {{f^2} + 1} \right)}^2}}} - {s_x}} \right){\bf{\hat x}}\\
 + \frac{1}{q}\left( {\frac{{2\sin (h)\left( { - s\left( {{f^2} - 1} \right)f' - f\left( {{f^2} + 1} \right){s_t}} \right) + 2f\left( {{f^2} + 1} \right)h'\cos (h)s}}{{{{\left( {{f^2} + 1} \right)}^2}}} - {s_y}} \right){\bf{\hat y}}\\
 + \frac{1}{q}\left( {\left( {\frac{2}{{{f^2} + 1}} - 1} \right){s_t} - {s_z} + \frac{{4ff's}}{{{{\left( {{f^2} + 1} \right)}^2}}}} \right){\bf{\hat z}}
\end{array}
\end{equation}
and
\begin{equation}\label{11}
\begin{array}{c}
{\bf{B}} = \frac{1}{q}\left( {\frac{{2\sin (h)\left( {s\left( {{f^2} - 1} \right)f' - f\left( {{f^2} + 1} \right){s_z}} \right) + \left( {{f^4} - 1} \right){s_y} - 2f\left( {{f^2} + 1} \right)h'\cos (h)s}}{{{{\left( {{f^2} + 1} \right)}^2}}}} \right){\bf{\hat x}}\\
 + \frac{1}{q}\left( {\frac{{2\cos (h)\left( {f\left( {{f^2} + 1} \right){s_z} - s\left( {{f^2} - 1} \right)f'} \right) - \left( {{f^4} - 1} \right){s_x} - 2f\left( {{f^2} + 1} \right)h'\sin (h)s}}{{{{\left( {{f^2} + 1} \right)}^2}}}} \right){\bf{\hat y}}\\
 + \frac{1}{q}\left( {\frac{{2f\left( {\sin (h){s_x} - \cos (h){s_y}} \right)}}{{{f^2} + 1}}} \right){\bf{\hat z}}
\end{array}
\end{equation}
Here, $U = {{{b_0}} \mathord{\left/ {\vphantom {{{b_0}} q}} \right. \kern-\nulldelimiterspace}q}$ is the electric potential,  ${\bf{A}} =  - \left( {{1 \mathord{\left/ {\vphantom {1 q}} \right.
 \kern-\nulldelimiterspace} q}} \right)\left( {{b_1}{\bf{\hat x}} + {b_2}{\bf{\hat y}} + {b_3}{\bf{\hat z}}} \right)$ is the magnetic vector potential and ${\bf{\hat x}}$, ${\bf{\hat y}}$, ${\bf{\hat z}}$ are the unit vectors along the x, y, z axes respectively.  It should also be noted that the choice of the minus sign in the definition of the magnetic vector potential is related to the form of the Dirac matrices ${\gamma ^\mu }$ used in this article. Furthermore, the prime denotes differentiation of the functions with respect to their arguments, e.g. $f' = {{df} \mathord{\left/ {\vphantom {{df} {d\left( {z - t} \right)}}} \right. \kern-\nulldelimiterspace} {d\left( {z - t} \right)}}$, while the subscript denotes partial differentiation of the functions with respect to the depicted variable, e.g. ${s_t} = {{\partial s} \mathord{\left/
 {\vphantom {{\partial s} {\partial t}}} \right.\kern-\nulldelimiterspace} {\partial t}}$.

An interesting characteristic of the solutions given by Eq. (8) is that they can become partially localized around the position $z = t$ , assuming that the function $f$ is not constant, e. g. $f$  is a Gaussian function of $z-t$. More details on localized solutions to the massless Dirac and Weyl equations can be found in \cite{4}.  

However, in the case that the functions $f,g$  are constant the electromagnetic fields given by equations (10), (11) are substantially simplified to the following form

\begin{equation}\label{12}
{\bf{E}} = \frac{{{c_0}h'}}{q}\left( { - \sin \left( h \right){\bf{\hat x}} + \cos \left( h \right){\bf{\hat y}}} \right)
\end{equation}

\begin{equation}\label{13}
{\bf{B}} = \frac{{{c_0}h'}}{q}\left( {\cos \left( h \right){\bf{\hat x}} - \sin \left( h \right){\bf{\hat y}}} \right)
\end{equation}

corresponding to a circularly polarized electromagnetic wave propagating along the z axis with Poynting vector 

\begin{equation}\label{14}
{\bf{S}} = \frac{1}{{4\pi }}{\bf{E}} \times {\bf{B}} = \frac{1}{{4\pi }}{\left( {\frac{{{c_0}h'}}{q}} \right)^2}{\bf{\hat z}}
\end{equation}

Here, ${c_0}$  is an arbitrary real constant. It is also worth mentioning that the spin of the particles, as calculated through the formulae \cite{2}
\begin{equation}\label{15}
{S_x} = \frac{i}{2}{\Psi ^\dag }{\gamma ^2}{\gamma ^3}\Psi  = 2\left( {{c_1} - {c_2}} \right)\left( {{c_1} + {c_2}} \right)f\cos \left( h \right)
\end{equation}

\begin{equation}\label{16}
{S_y} = \frac{i}{2}{\Psi ^\dag }{\gamma ^3}{\gamma ^1}\Psi  = 2\left( {{c_1} - {c_2}} \right)\left( {{c_1} + {c_2}} \right)f\sin \left( h \right)
\end{equation}

\begin{equation}\label{17}
{S_z} = \frac{i}{2}{\Psi ^\dag }{\gamma ^1}{\gamma ^2}\Psi  = \left( {{c_1} - {c_2}} \right)\left( {{c_1} + {c_2}} \right)\left( { - 1 + {f^2}} \right)
\end{equation}
 
rotates on the x-y plane in synchronization with the circularly polarized electromagnetic wave given by equations (12), (13). 

This becomes clearer if we suppose that the arbitrary real function $h$  takes the form $h=k(z-t)$, where $k$ is an arbitrary real constant. In this case, equations (12)-(16) become 	

\begin{equation}\label{18}
{\bf{E}} = \frac{{{c_0}k}}{q}\left( { - \sin \left( {k\left( {z - t} \right)} \right){\bf{\hat x}} + \cos \left( {k\left( {z - t} \right)} \right){\bf{\hat y}}} \right)
\end{equation}

\begin{equation}\label{19}
{\bf{B}} = \frac{{{c_0}k}}{q}\left( {\cos \left( {k\left( {z - t} \right)} \right){\bf{\hat x}} - \sin \left( {k\left( {z - t} \right)} \right){\bf{\hat y}}} \right)
\end{equation}

\begin{equation}\label{20}
{\bf{S}} = \frac{1}{{4\pi }}{\bf{E}} \times {\bf{B}} = \frac{1}{{4\pi }}{\left( {\frac{{{c_0}k}}{q}} \right)^2}{\bf{\hat z}}
\end{equation}

and

\begin{equation}\label{21}
{S_x} = \frac{i}{2}{\Psi ^\dag }{\gamma ^2}{\gamma ^3}\Psi  = 2\left( {{c_1} - {c_2}} \right)\left( {{c_1} + {c_2}} \right){f_0}\cos \left( {k\left( {z - t} \right)} \right)
\end{equation}

\begin{equation}\label{22}
{S_y} = \frac{i}{2}{\Psi ^\dag }{\gamma ^3}{\gamma ^1}\Psi  = 2\left( {{c_1} - {c_2}} \right)\left( {{c_1} + {c_2}} \right){f_0}\sin \left( {k\left( {z - t} \right)} \right)
\end{equation}

\begin{equation}\label{23}
{S_z} = \frac{i}{2}{\Psi ^\dag }{\gamma ^1}{\gamma ^2}\Psi  = \left( {{c_1} - {c_2}} \right)\left( {{c_1} + {c_2}} \right)\left( { - 1 + f_0^2} \right)
\end{equation}

where we have also replaced $f$  by $f_0$, since we have supposed that the function $f$ is constant. Here, it should also be mentioned that the total spin of the particles, given by the formula

\begin{equation}\label{24}
S = {\left( {S_x^2 + S_y^2 + S_z^2} \right)^{1/2}} = \left| {c_1^2 - c_2^2} \right|\left( {1 + f_0^2} \right)
\end{equation}

can take any desired value through an appropriate choice of the constants $c_1,c_2,f_0$. Thus, the spin of the particles can either be set equal to ½ corresponding to single particles, or to any other desired value, corresponding to systems of particles with any combination of the relative orientation of their spins.

From the above analysis it is evident that the x and y components of the spin of the particles rotate with the same frequency as the circularly polarized electromagnetic wave given by equations (18), (19). Consequently, the synchronization of the spin of the particles with the incident circularly polarized electromagnetic wave can be considered as a key feature of the degenerate solutions given by Eq. (7). Furthermore, since the particles are massless, this result may have applications to graphene, which supports massless Dirac fermions \cite{5,6,7,8,9}. In more detail, our results indicate that in the presence of a circularly polarized electromagnetic wave, e.g. a laser beam, the spin of the massless Dirac fermions in graphene rotates in synchronization with the incident electromagnetic wave. In addition, the rotation of the spin does not change in the presence of the more general electromagnetic fields given by equations (10), (11), or even in the absence of electromagnetic 4-potentials and fields, as it can be easily deduced by setting the arbitrary function $s$ equal to zero in Eq. (8). 

Regarding the potential practical applications of the above results, the fact that the degenerate spinors given by Eq. (7) are also solutions to the massless Dirac equation even for zero electromagnetic 4-potential and field could be used for the development of optical memories based on massless charged particles in graphene and other materials. In Figure 1 we describe an experimental setup proposed for this purpose, which is analogous to the setups that have been used for the development of three-dimensional optical memories \cite{12} and in two-photon microscopy \cite{13}.

\begin{figure}[h] % [h] stands for 'here'
    \centering
    \includegraphics[width=0.6\textwidth]{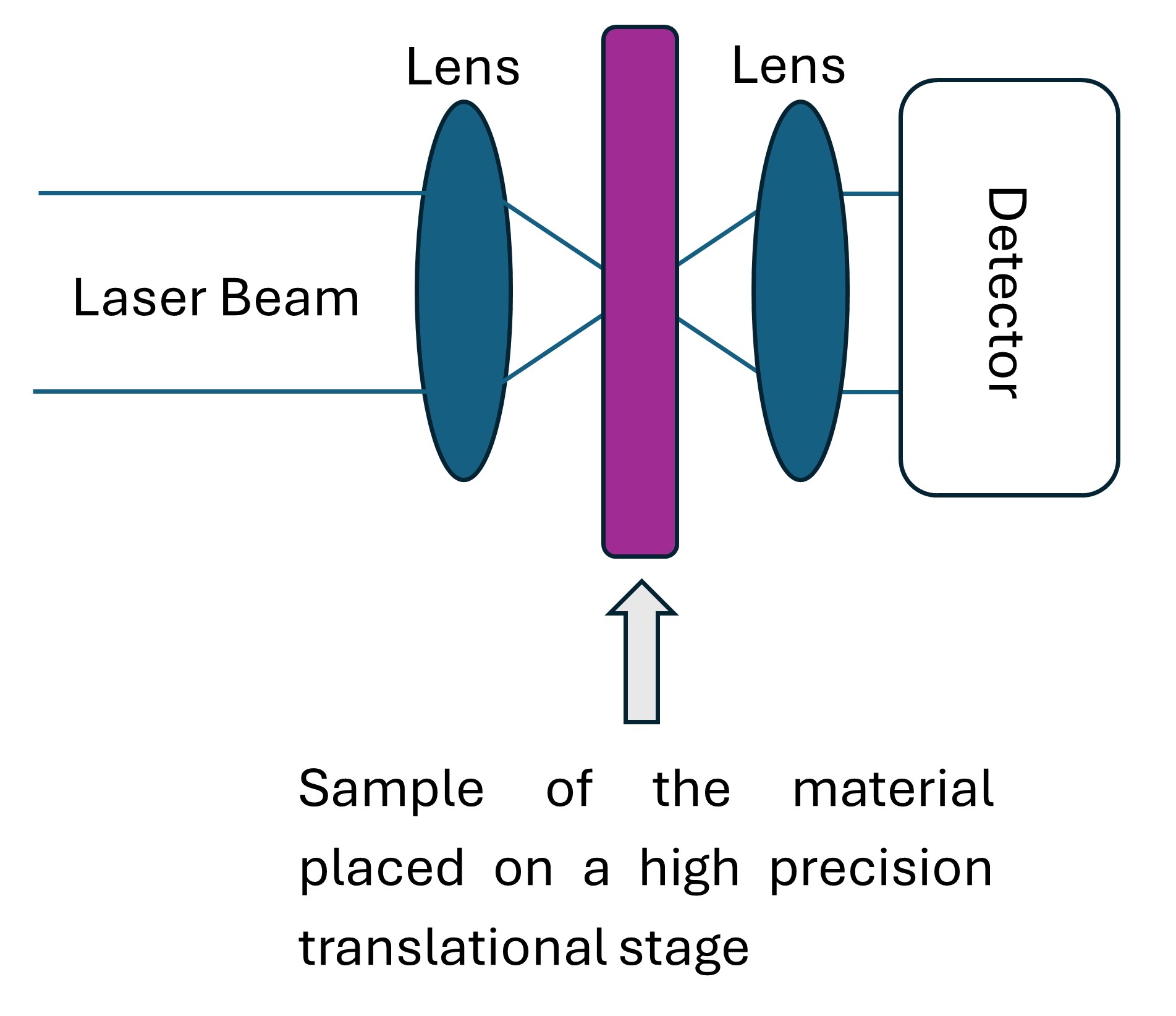} % Do not include .jpg or .png
    \caption{Proposed experimental setup for the development of optical memories in graphene and other materials supporting massless Dirac fermions.}
\end{figure}

\begin{figure}[h] % [h] stands for 'here'
    \centering
    \includegraphics[width=0.3\textwidth]{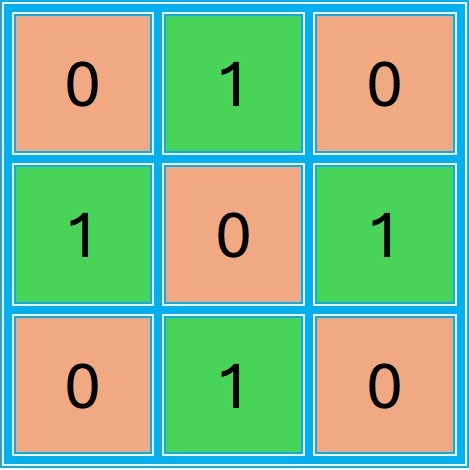} % Do not include .jpg or .png
    \caption{A sequence of bits recorded on a material supporting massless Dirac fermions. The blue line corresponds to the insulating layer used to suppress the diffusion of the carriers between the bits.}
\end{figure}

In more detail, a thin sample of the material is placed on a high-precision micrometric stage and moved transversely in the focal region of a laser beam. Dividing the material into small pixels with an insulating layer between them to avoid diffusion of the carriers, information could be written on each pixel using a circularly polarized laser beam at frequency $\omega$. Specifically, the spin of the particles in the pixels irradiated by the beam will rotate with frequency $\omega$. Afterwards, the information could be retrieved using a laser beam of the same frequency. In more detail, the pixels that have been irradiated with the writing beam will be in a degenerate state and will not interact with the reading beam, contrary to those that have not been irradiated. In this way, one could write and read a sequence of binary data (ones and zeros) on a thin layer of a material supporting massless charged particles, as shown in Figure 2.

In conclusion, we have presented a novel class of degenerate solutions to the massless Dirac equation and discussed their potential applications on the development of optical memories using material supporting massless Dirac fermions, such as graphene.

\end{document}